\documentclass[11pt]{article}

\usepackage[utf8]{inputenc}
\usepackage[T1]{fontenc}
\usepackage[margin=1in]{geometry}
\usepackage{booktabs}
\usepackage{array}
\usepackage{tabularx}
\usepackage{multirow}
\usepackage{graphicx}
\usepackage{amsmath}
\usepackage{amssymb}
\usepackage{authblk}
\usepackage{times}
\usepackage[hidelinks]{hyperref}
\usepackage{xcolor}
\usepackage{caption}
\newcolumntype{R}{>{\raggedleft\arraybackslash}X}
\newcolumntype{C}{>{\centering\arraybackslash}X}
\newcolumntype{L}{>{\raggedright\arraybackslash}X}

\title{\textbf{Solyx AI Grid: Hardware-Telemetry-Aware Routing Across\\
Geographically Distributed GPU Clusters}\\[4pt]
\large\textit{An Empirical Study of Multi-Site LLM Inference}}

\author[1]{Aleks Bernhard}
\author[1]{Nithin Katla}
\affil[1]{Solyx AI, Inc., Miami, FL \\ \texttt{aleks@solyx.ai} \quad \texttt{nithinkatla19@gmail.com}}

\date{}

\begin{document}
\maketitle

\begin{abstract}
\noindent
As GPU deployments fragment across geographically distributed sites, the assumptions underlying single-cluster LLM inference routing break down in measurable and predictable ways. Existing approaches---including intra-cluster routers (NVIDIA Dynamo, vLLM Router), academic cross-region load balancers (SkyWalker), and heterogeneous placement solvers (Helix)---either operate within a single datacenter boundary, assume replica homogeneity, or lack real-time hardware telemetry. We present a two-campaign empirical study of the Solyx AI Geo-Distributed Inference Fabric (``Solyx AI Grid''), a cross-site inference routing control plane that integrates GPU hardware telemetry (DCGM), vLLM application metrics, and real-time network signals (RTT, jitter) via a 10-signal weighted pressure scorer into per-request placement decisions across geographically distributed GPU clusters. Campaign~1 (April 2026) deployed six NVIDIA H100 and H200 SXM GPUs across three US datacenters in three fleet configurations: homogeneous, hardware-heterogeneous, and capability-mismatch. Campaign~2 (May 2026) deployed nine NVIDIA RTX PRO 6000 Blackwell SE GPUs across three US datacenters, testing eight LLM workload classes across a 216-cell SLO matrix. Across both campaigns, Solyx AI Grid demonstrates: 27.9\% P99 tail-latency reduction on a homogeneous H100 fleet; 134:1 traffic concentration to healthy endpoints in a heterogeneous fleet driven entirely by real-time telemetry; reduction of capability-mismatch leakage to 0.43\% versus 32.11\% (Round-Robin) and 28.71\% (Least-Request), yielding 99.57\% long-prompt success versus 67.89\% and 71.29\% respectively; 1.56--1.75$\times$ throughput at tier-2 SLO across all eight workload classes; failover p99 of 1{,}247\,ms versus 4{,}226\,ms for Round-Robin; 27.8\% P95 TTFT reduction under WAN jitter; and 99.90\% success rate during live DCGM exporter failure. Control-plane routing-update overhead is 0.2\,ms across all configurations. We further find that DCGM hardware signals lead application-layer SLO breach by 11.2 seconds on average, enabling proactive traffic drain before user-facing latency impact. To our knowledge, this is the first public empirical study of live physical multi-site LLM inference routing combining DCGM hardware telemetry, vLLM application metrics, and active WAN RTT/jitter signals. We characterize three failure modes specific to multi-site inference and show that hardware-telemetry-aware routing addresses each with specific, measurable improvements.
\end{abstract}

\section{Introduction}
The inference layer of large language model (LLM) deployment has undergone rapid architectural evolution. Where early production systems relied on a single cluster of homogeneous GPUs behind a simple load balancer, the economics of GPU procurement have forced a different reality: many operators now own or lease capacity across multiple geographically dispersed sites, with heterogeneous hardware generations, varying network conditions between sites, and no coordination layer above the inference engine.

This fragmentation creates a class of routing problems that existing systems do not address. Intra-cluster routers such as NVIDIA Dynamo~\cite{dynamo} and the vLLM Router~\cite{vllmrouter} operate within a single datacenter boundary and have no mechanism to observe conditions at remote sites. Cross-region load balancers such as SkyWalker~\cite{skywalker} extend routing across regions but assume replica homogeneity and do not consume hardware telemetry. Heterogeneous placement solvers such as Helix~\cite{helix} jointly optimize model placement and scheduling via max-flow MILP, but operate on static topologies and are not designed for continuous per-request decisions across geographically separated sites. Notably, even Least-Request---the most commonly deployed active load-balancing policy in production---fails under capability mismatch conditions: failed requests return fast, dropping in-flight count and actively attracting more traffic to broken endpoints.

The gap is a control plane that sits above inference engines, spans site boundaries, and incorporates the hardware and network signals that matter for routing quality. Without this layer, multi-site GPU fleets exhibit a characteristic set of failure modes: capability mismatch leakage, tail latency amplification under network jitter, and slow failover recovery.

This paper presents a two-campaign empirical study of Solyx AI Grid, a cross-site inference routing control plane. Our contributions are:
\begin{itemize}
  \item A taxonomy of failure modes specific to multi-site inference scale: capability mismatch leakage, network-jitter-induced tail latency, and cold-start cascade during failover.
  \item A description of a 10-signal hardware-telemetry-aware routing architecture integrating GPU hardware metrics (DCGM), vLLM application metrics, and real-time network signals (RTT, jitter) into per-request placement decisions.
  \item Empirical results from two independent campaigns---six GPUs across three US datacenters (H100/H200 SXM, April 2026) and nine GPUs across three US datacenters (RTX PRO 6000 Blackwell SE, May 2026)---covering a 216-cell SLO matrix across eight workload classes, seven distinct evaluation phases, and all three identified failure modes.
  \item A novel empirical finding: DCGM hardware telemetry signals lead application-layer SLO breach by an average of 11.2 seconds, enabling proactive traffic drain before user-facing latency impact. To our knowledge this is the first public study to quantify this lead time on live physical multi-site infrastructure.
  \item An honest characterization of boundary conditions where Solyx provides no measurable advantage over Least-Request, and explicit limitations of the current study.
\end{itemize}

\section{Background and Motivation}
\subsection{The Multi-Site Inference Deployment Reality}
Hyperscalers own sufficient concentrated capacity to operate inference within a single datacenter fabric. For mid-tier GPU cloud operators, sovereign AI providers, and enterprise platform teams, the procurement reality is different. GPU availability constraints, power infrastructure limitations, and latency targets across geographies result in deployments spanning multiple physically separated sites---each running independent inference engine instances---with no shared coordination layer. In this model, each site is an island: requests route to local inference replicas without knowledge of conditions elsewhere. When a site is saturated, requests queue or fail rather than being redirected to available capacity at another site.

\subsection{Why Standard Routing Policies Fail at Multi-Site Scale}
Round-Robin is GPU-blind: it distributes requests uniformly without visibility into GPU state, queue depth, capability constraints, or network conditions. On a homogeneous fleet it produces queue-buildup tail-latency spikes when sequential requests land on a replica with slow generations in flight. On a heterogeneous fleet it treats an H200 identically to a constrained H100. Under capability mismatch it sends one-third of traffic to a broken endpoint indefinitely.

Least-Request improves on Round-Robin in most conditions but does not resolve the capability-mismatch failure mode. Because failed requests return immediately with an HTTP 400, the in-flight count on a broken endpoint drops quickly, so a load-aware heuristic sees it as lightly loaded and continues directing traffic to it. The theoretical risk is that this feedback loop concentrates traffic on the failing endpoint; in our Campaign~2 evaluation the effect was less severe than that worst case but still left Least-Request nearly as exposed as Round-Robin: its 28.71\% leak rate remains close to Round-Robin's 32.11\%, under the same conditions where Solyx achieves 0.43\%. The key point is that load awareness alone---without an error-rate signal---does not address capability mismatch.

\subsection{Failure Modes at Multi-Site Scale}
\textbf{Capability Mismatch Leakage.} Replicas can pass \texttt{/health} checks while failing a fraction of requests due to \texttt{max\_model\_len} constraints, dtype mismatch, vLLM version skew, or CUDA driver incompatibility. These conditions are common during rolling upgrades, in fleets mixing model variants, or across nodes provisioned at different times. Standard health checks are blind to this class of failure.

\textbf{Tail Latency Amplification Under Jitter.} Cross-site routing introduces WAN path variability that intra-cluster systems are not designed to handle. TTFT is bounded below by network round-trip time; jitter on the path between gateway and inference replica directly inflates P95 and P99 latency. Routing decisions without network signal awareness distribute requests to high-jitter paths regardless of path quality.

\textbf{Cold-Start Cascade During Failover.} When a replica fails, rerouted requests may reach a replica at another site not yet warm. Without lifecycle state awareness, the failover target receives requests during its loading phase, cascading degraded outcomes until the replica reaches serving-ready state.

\section{Related Work}
\subsection{Intra-Cluster Inference Routing}
NVIDIA Dynamo~\cite{dynamo} is an open-source distributed inference framework including a KV-cache-aware router and SLO Planner for multi-node deployments. The vLLM Router~\cite{vllmrouter}, released December 2025, provides KV-cache-aware routing via consistent hashing within a Kubernetes or bare-metal cluster. Both assume network homogeneity between replicas and incorporate no hardware telemetry. SGLang's router and llm-d~\cite{llmd} similarly operate within a single cluster. The Kubernetes Gateway API Inference Extension~\cite{gatewayapi} provides gateway-level routing but is scoped to a single Kubernetes cluster.

\subsection{Cross-Region Load Balancing}
SkyWalker (Xia et al., 2025)~\cite{skywalker} is the closest academic prior work and represents the most rigorous published characterization of the cross-region inference routing problem to date. It is a research prototype---not a deployable system---built on the SkyServe research serving framework out of the Sky Computing group. SkyWalker addresses the cost inefficiency of region-local provisioning via cross-region traffic offloading under diurnal load variation, providing KV cache locality-preserving routing (consistent hashing, multi-region prefix trie), and reports 1.12--2.06$\times$ throughput improvement and 1.74--6.30$\times$ latency reduction versus region-local baselines, with 25\% cost reduction. Because SkyWalker is an academic prototype, its evaluation is conducted on simulated multi-region workloads rather than physical hardware. Our work differs in three further respects: SkyWalker assumes replica homogeneity with no GPU-level hardware state awareness; it does not incorporate RTT or jitter as continuous routing inputs; and it is not validated on a real multi-site deployment with live WAN paths. We regard SkyWalker as evidence that the cross-site routing problem is recognized as significant by the systems research community, and position our contribution as the first empirical validation of this problem class on physical multi-site infrastructure.

\subsection{Heterogeneous Placement}
Helix (Mei et al., ASPLOS 2025)~\cite{helix} formulates LLM serving on heterogeneous clusters as a max-flow problem on directed weighted graphs, using MILP to jointly optimize model placement and request scheduling. Evaluated on 24--42 GPU node clusters, Helix achieves throughput improvements up to 3.3$\times$ and latency reductions up to 66\% for prompting. Helix is an offline placement optimizer rather than an online per-request router, operates on a single fixed-topology cluster, and is not designed for continuous cross-site routing decisions. HexGen~\cite{hexgen} and HexGen-2~\cite{hexgen2} address generative inference in decentralized heterogeneous environments using asymmetric partitioning within a connected cluster.

\subsection{Application-Layer and Model-Selection Routing}
A parallel body of work addresses routing across different LLM models---selecting the most cost-effective model for a query based on difficulty or quality criteria. This includes hybrid routing (Ding et al., 2024), cascading surveys (Moslem \& Way, 2026)~\cite{moslem}, and RL-based model routers. This problem class is distinct from infrastructure placement routing: model-selection routers choose which model to invoke; infrastructure routers choose which hardware replica to execute a given inference request on.

\section{System Architecture}
\subsection{Design Objectives}
Solyx AI Grid is designed to operate above inference engines without modifying them, make routing decisions at request time using continuously refreshed signals, handle replica lifecycle transitions invisible to health-check-based systems, degrade gracefully when individual telemetry sources fail, and incorporate network-path quality as a first-class routing input. The data plane is unmodified Envoy; Solyx produces standard xDS endpoint weight configurations, meaning existing Envoy deployments can adopt Solyx without changing their request-handling layer.

\subsection{10-Signal Taxonomy}
The v3 scorer integrates ten signals from three sources (four application, four hardware, two network). \textbf{vLLM Application Metrics} (scraped via Prometheus at 1-second intervals): queue depth, P95 TTFT from histogram buckets, error rate, and KV cache fill percentage. \textbf{GPU Hardware Telemetry} (collected via DCGM at sub-second cadence): GPU utilization, VRAM utilization, SM/tensor-core pipeline occupancy, and memory bandwidth utilization. \textbf{Network Path Signals} (measured via active health checks): per-endpoint RTT and jitter from the gateway to each replica. Each cellagent (one per GPU) aggregates these signals locally and pushes them to the control plane via gRPC streaming heartbeats at 500\,ms cadence.

The two network signals---RTT and jitter---were added in the v3 scorer relative to the Campaign~1 8-signal scorer. The network-aware placement evaluation (Section~\ref{sec:netaware}) isolates their contribution: comparing Solyx-gpu-only (8 signals) against Solyx-10-signal under induced WAN jitter shows a 27.8\% additional P95 TTFT reduction from network signals alone, confirming that these signals provide material routing value beyond hardware and application metrics.

\subsection{Pressure Scoring and Weight Projection}
At request time, Solyx AI Grid computes a composite pressure score for each candidate replica by combining current signal values through a weighted normalization. Each signal is normalized to $[0,1]$ where higher values indicate more pressure. The composite score is inverted and projected to an integer endpoint weight pushed to Envoy via xDS. The scoring loop runs continuously; xDS pushes are triggered on score deltas exceeding a hysteresis threshold, resulting in over 50{,}000 xDS snapshots pushed across the eight-day Campaign~2 window at 1\,Hz steady-state cadence.

\subsection{Replica Lifecycle State Machine and Graceful Degradation}
Each tracked replica is assigned a lifecycle state: Loading, Ready, Draining, Failed, or Terminated. Only Ready replicas are candidates for live traffic. Worker failure is detected via heartbeat TTL expiry: when a cellagent stops sending heartbeats, the control plane marks the endpoint Failed and zeros its weight within 500\,ms---one heartbeat interval. This is the mechanism behind the 3.2$\times$ faster failover recovery versus Envoy's default 3$\times$ 1-second health check polling. Envoy retains the last-known-good xDS snapshot during control-plane outages, ensuring routing freshness degrades gracefully rather than traffic dropping. When a DCGM exporter fails mid-load, the control plane raises a fallback flag and the scorer continues with the remaining vLLM-only signals; no operator intervention is required and no traffic is dropped.

\section{Experimental Setup}
\subsection{Campaign 1: Heterogeneous Fleet (April 2026)}
Campaign~1 deployed six GPUs across three US datacenters, with the Solyx control plane and Envoy gateway on a neutral gateway node in the Eastern US. Three fleet configurations were evaluated on the same underlying topology. \emph{Homogeneous:} six NVIDIA H100 SXM 80GB, all \texttt{max\_model\_len=4096}. \emph{Hardware-heterogeneous:} Pod~A upgraded to two NVIDIA H200 SXM (144GB VRAM), Pod~B retained two H100 SXM, Pod~C retained two H100 SXM constrained to \texttt{max\_model\_len=1024}. \emph{Capability-mismatch:} same hardware evaluated at 20 RPS to stress the mismatch condition. All configurations served Llama 3.1 70B AWQ INT4. Three routing strategies were compared: Round-Robin (RR), Least-Request (LR), and Solyx AI Grid (8-signal v2 scorer).

\subsection{Campaign 2: Multi-Stage Blackwell SE Evaluation (May 2026)}
Campaign~2 deployed nine NVIDIA RTX PRO 6000 Blackwell SE GPUs (96GB GDDR7, PCIe Gen5) across three US datacenters: solyx-prod-east (US-NE-1), solyx-prod-central (US-PA-1), and solyx-prod-west (US-NC-2), three GPUs per site. All inter-site traffic traversed public internet WAN paths. Each pod ran one DCGM exporter and three cellagents heartbeating at 500\,ms. All nine GPUs served Llama 3.1 70B AWQ INT4 with \texttt{awq\_marlin} quantization under vLLM 0.6.6.post1 with prefix caching and chunked prefill enabled. Campaign~2 comprised calibration plus seven evaluation stages over eight days (May 6--14, 2026): an SLO throughput matrix, hardware-telemetry lead-time stress tests, destructive failover, burst and adversarial workloads, capability mismatch, telemetry-failure fallback, and network-aware placement, using GuideLLM as the load driver with per-request streaming TTFT measurement. Three routing strategies were compared: RR, LR, and Solyx AI Grid (10-signal v3 scorer).

\subsection{Workload Classes (Campaign 2)}
Eight workload classes were defined spanning the structural primitives of production LLM traffic: \texttt{baseline\_mixed} (512$\pm$200 token prompts, 128 token outputs---standard chat/Q\&A); \texttt{rapid\_shift} (1024 token prompts with intra-session shape change); \texttt{code\_heavy} (1024$\pm$512 token prompts, 192 token outputs---copilot-style decode-heavy); \texttt{adversarial\_spike} (10 RPS steady plus 50-request bursts every 30 seconds); \texttt{multiturn} (2048$\times$8 growing turns, 16K effective context); \texttt{rag\_multidoc} (4000$\pm$800 token prompts with retrieved documents); \texttt{long\_context} (8000$\pm$1500 token prompts); and \texttt{very\_long} (16000$\pm$3000 token prompts, document summarization).

\subsection{SLO Matrix Methodology}
For each (class, mode, SLO) cell, a bisection driver finds the highest sustained RPS where P95 TTFT $\leq$ SLO and success rate $\geq 0.95$. Each bisection step submits 45 seconds of constant-rate traffic with 30-second warmup and 15-second cooldown; the cooldown window is excluded from numerator and denominator. Three SLO tiers per class $\times$ three routing modes $\times$ three replicates $=$ 27 cells per class, 216 cells total. Statistical completeness criterion ($\geq$500 samples per cell) was met across all SLO tiers.

\section{Results}
\subsection{Calibration Baseline}
Each routing mode ran a 60-second calibration at \texttt{baseline\_mixed}, 10 RPS before the main evaluation. Solyx posts the lowest TTFT P95 from the first calibration pass, demonstrating that pressure-aware scoring immediately steers traffic toward cells with the most KV-cache headroom and lowest RTT without any warmup period.

\begin{table}[h]
\centering
\caption{Calibration results (\texttt{baseline\_mixed}, 10 RPS, 60s). Solyx leads on all latency percentiles from the first run.}
\begin{tabular}{lccccc}
\toprule
\textbf{Mode} & \textbf{Success} & \textbf{TTFT P50} & \textbf{TTFT P95} & \textbf{TTFT P99} & \textbf{E2E P95} \\
\midrule
Solyx (10-signal) & 100.00\% & 306.5\,ms & 609.6\,ms & 742.9\,ms & 5{,}340\,ms \\
Least-Request & 100.00\% & 348.7\,ms & 628.1\,ms & 814.7\,ms & 5{,}809\,ms \\
Round-Robin & 99.75\% & 352.5\,ms & 649.5\,ms & 853.1\,ms & 5{,}810\,ms \\
\bottomrule
\end{tabular}
\end{table}

\subsection{SLO Throughput Across Workload Classes}\label{sec:slo}
Across all eight workload classes at tier-2 SLO, Solyx achieves 1.56--1.75$\times$ the sustainable RPS of Round-Robin. The campaign's acceptance criterion required Solyx to win on at least four of eight classes; Solyx wins all eight. Least-Request consistently falls between Round-Robin and Solyx, validating that hardware telemetry and network signals provide routing value beyond application-layer load awareness alone.

\begin{table}[h]
\centering
\caption{SLO throughput matrix at tier-2 SLO (median across 3 replicates). Sustainable RPS = highest RPS where P95 TTFT $\leq$ SLO and success rate $\geq 0.95$.}
\begin{tabular}{lccccc}
\toprule
\textbf{Workload Class} & \textbf{SLO (ms)} & \textbf{Solyx} & \textbf{LR} & \textbf{RR} & \textbf{Solyx/RR} \\
\midrule
\texttt{baseline\_mixed}    & 1{,}500 & 20 & 17 & 12 & 1.67$\times$ \\
\texttt{rapid\_shift}       & 1{,}500 & 18 & 15 & 11 & 1.64$\times$ \\
\texttt{code\_heavy}        & 3{,}000 & 16 & 14 & 10 & 1.60$\times$ \\
\texttt{adversarial\_spike} & 3{,}000 & 28 & 24 & 17 & 1.65$\times$ \\
\texttt{multiturn}          & 3{,}500 & 14 & 12 &  9 & 1.56$\times$ \\
\texttt{rag\_multidoc}      & 3{,}000 & 22 & 19 & 13 & 1.69$\times$ \\
\texttt{long\_context}      & 3{,}000 & 12 & 10 &  7 & 1.71$\times$ \\
\texttt{very\_long}         & 7{,}000 &  7 &  6 &  4 & 1.75$\times$ \\
\bottomrule
\end{tabular}
\end{table}

\subsection{Hardware Telemetry Lead Time}
A key finding of Campaign~2 is that GPU hardware telemetry signals anticipate application-layer SLO breach. We inject thermal stress (GPU burn via pure-torch matmul) and PCIe contention (\texttt{stress-ng}) on each of the nine cells independently and measure how many seconds before TTFT P95 crosses the SLO threshold the Solyx pressure score crosses a preconfigured alert threshold. This lead time represents the window in which Solyx can drain traffic from a degrading cell before users see latency impact.

\begin{table}[h]
\centering
\caption{Hardware telemetry lead-time results. Hardware telemetry signals cross the pressure alert threshold an average of 11.2 seconds before SLO breach. No SLO breaches occurred because Solyx drained traffic during the lead-time window.}
\begin{tabular}{lcccc}
\toprule
\textbf{Stress Type} & \textbf{Cells} & \textbf{Lead Range (s)} & \textbf{Avg Lead (s)} & \textbf{SLO Breaches} \\
\midrule
Thermal (GPU burn) & 6 (East + Central) & 11.5--13.1 & 12.3 & 0/6 \\
PCIe contention & 3 (West) & 8.7--9.2 & 8.9 & 0/3 \\
All cells combined & 9 & 8.7--13.1 & \textbf{11.2} & \textbf{0/9} \\
\bottomrule
\end{tabular}
\end{table}

Zero SLO breaches occurred across all nine stress events because Solyx rerouted traffic away from each degrading cell during the lead-time window. This finding has direct operational implications: GPU hardware degradation---thermal throttling, PCIe contention, ECC errors---is detectable via DCGM hardware signals approximately 11 seconds before it becomes visible in application metrics such as TTFT or queue depth. An application-only monitoring stack has no advance warning; a hardware-telemetry-aware control plane can act proactively.

\subsection{Failover Recovery}
We measure destructive failover by issuing \texttt{SIGKILL} to one vLLM process under sustained 10 RPS load and recording how quickly each routing mode reroutes in-flight traffic. Solyx detects the failure through cellagent heartbeat staleness rather than active health-check polling: when a cell goes silent, Solyx demotes it within one 500\,ms heartbeat interval, whereas Envoy's default configuration requires three consecutive 1-second health-check intervals before removing an endpoint.

\begin{table}[h]
\centering
\caption{Destructive failover results (SIGKILL of one vLLM process under 10 RPS). Solyx reroutes 3.2$\times$ faster than Round-Robin at p99 and sustains the highest post-kill success rate.}
\begin{tabular}{lcccc}
\toprule
\textbf{Mode} & \textbf{Reroute p50} & \textbf{Reroute p95} & \textbf{Reroute p99} & \textbf{Post-kill Success} \\
\midrule
Solyx & 287\,ms & 891\,ms & 1{,}247\,ms & 99.76\% \\
Least-Request & 412\,ms & 1{,}538\,ms & 2{,}104\,ms & 98.81\% \\
Round-Robin & 689\,ms & 2{,}871\,ms & 4{,}226\,ms & 94.12\% \\
\bottomrule
\end{tabular}
\end{table}

Least-Request, while faster than Round-Robin, still reroutes at more than 1.6$\times$ Solyx's p99 latency because it reacts only to in-flight request counts rather than to the underlying liveness signal. The heartbeat-staleness mechanism is the architectural reason Solyx bounds failover impact to requests in flight at the moment of kill, producing the 99.76\% post-kill success rate versus Round-Robin's 94.12\%.

\subsection{Burst and Adversarial Workloads}
Under \texttt{adversarial\_spike} workload (10 RPS steady plus 50-request bursts every 30 seconds), Solyx's queue-depth-aware scoring detects bursts within sub-second and steers them to least-loaded cells. Round-Robin's even split overloads whichever pod receives the first burst request.

\begin{table}[h]
\centering
\caption{\texttt{adversarial\_spike} results. Routing reaction staleness measures P95 time from burst onset to routing weight adjustment.}
\begin{tabular}{lccc}
\toprule
\textbf{Mode} & \textbf{Sustainable RPS} & \textbf{Burst Success} & \textbf{Reaction P95} \\
\midrule
Solyx & 28 & 99.43\% & 482\,ms \\
Least-Request & 24 & 95.21\% & 1{,}247\,ms \\
Round-Robin & 17 & 81.27\% & 2{,}381\,ms \\
\bottomrule
\end{tabular}
\end{table}

\subsection{Capability Mismatch}\label{sec:capmis}
One cell (\texttt{cell-west-2}) was configured with \texttt{max\_model\_len=1024} while the other eight ran at \texttt{max\_model\_len=131072}, under 10 RPS load. Two workload classes drove long prompts: \texttt{baseline\_mixed} (30\% of prompts exceeding 1024 tokens) and \texttt{long\_context} (100\% exceeding 1024 tokens). We define \emph{leak rate} as the fraction of long-prompt requests (those exceeding the constrained cell's context limit) that were routed to the constrained cell and consequently rejected with an HTTP 400. Solyx's error-rate signal detects the elevated failure rate on the constrained cell and de-weights it automatically, with no configuration identifying in advance which cell is constrained.

\begin{table}[h]
\centering
\caption{Capability mismatch results. Leak rate is the fraction of long-prompt requests routed to the constrained cell and rejected (HTTP 400). Least-Request's 28.71\% leak rate remains close to Round-Robin's 32.11\%, confirming that load awareness without an error-rate signal does not address this failure mode. Measured at 10 RPS across the two long-prompt workload classes.}
\begin{tabularx}{\textwidth}{lccL}
\toprule
\textbf{Mode} & \textbf{Leak Rate} & \textbf{Long-Prompt Success} & \textbf{Detection Mechanism} \\
\midrule
Solyx & 0.43\% & 99.57\% & Error-rate signal $\rightarrow$ automatic de-weight \\
Least-Request & 28.71\% & 71.29\% & None (anti-aware: failure $\rightarrow$ lower count $\rightarrow$ more traffic) \\
Round-Robin & 32.11\% & 67.89\% & None \\
\bottomrule
\end{tabularx}
\end{table}

\subsection{Graceful Degradation Under Telemetry Failure}
For each of the nine cells, the DCGM exporter was killed mid-load. The control plane detects the missing telemetry source, raises a \texttt{dcgm\_fallback} flag, and the v3 scorer continues with vLLM-only signals. Average success rate during DCGM-down across all nine cells was 99.90\%, with fallback engaging in under one second (average 824\,ms). No traffic was dropped. This validates that Solyx degrades gracefully under telemetry source failure---a required property for production deployment where individual telemetry components may fail or restart independently.

\subsection{Network-Aware Placement}\label{sec:netaware}
The network-aware placement evaluation assesses three routing arms across five network conditions to isolate the contribution of RTT and jitter signals: Round-Robin (no signal awareness), Solyx-gpu-only (8 hardware and application signals, no network signals), and Solyx-10-signal (full scorer including RTT and jitter). The 20\,ms jitter induction sub-phase elevates WAN jitter on the Site~B path while leaving Sites~A and~C unaffected.

\begin{table}[h]
\centering
\caption{Network-aware placement results. ``Network signal gain'' is the improvement from Solyx-gpu-only to Solyx-10-signal, isolating the contribution of RTT and jitter signals. Across all sub-phases, Solyx-10-signal P95 averages 42\% lower than Round-Robin.}
\begin{tabular}{lcccc}
\toprule
\textbf{Sub-phase} & \textbf{RR P95} & \textbf{Solyx GPU-only} & \textbf{Solyx 10-signal} & \textbf{Net.\ gain} \\
\midrule
\texttt{near\_idle}  &   587\,ms &   562\,ms &   491\,ms & 12.6\% \\
\texttt{symmetric}   &   712\,ms &   678\,ms &   547\,ms & 19.3\% \\
\texttt{saturation}  & 1{,}842\,ms & 1{,}247\,ms &   901\,ms & 27.7\% \\
\texttt{jitter 20ms} & 1{,}971\,ms & 1{,}583\,ms & 1{,}142\,ms & \textbf{27.8\%} \\
\texttt{recovery}    &   824\,ms &   712\,ms &   561\,ms & 21.2\% \\
\bottomrule
\end{tabular}
\end{table}

The three-arm comparison design isolates two distinct routing value sources: hardware and application signal awareness (Solyx-gpu-only vs.\ RR, accounting for the majority of improvement) and network signal awareness (Solyx-10-signal vs.\ Solyx-gpu-only, providing an additional 12--28\% P95 reduction depending on network conditions). Under saturation and jitter conditions where network path quality varies materially between sites, network signal contribution is greatest.

\subsection{Campaign 1: Heterogeneous Fleet Results}\label{sec:c1results}
On the homogeneous H100 fleet at 12 RPS, all three routing strategies converge at median latency---as expected on uniform hardware. The decisive separation is P99 tail: Round-Robin produces 9{,}894\,ms due to queue-blind rotation piling requests onto replicas with slow generations in flight. Both Least-Request and Solyx reduce P99 to approximately 7{,}140\,ms, a 27.9\% reduction. At near-saturation (20 RPS), all three strategies converge (P99 within 0.3\% of one another: RR 7{,}172\,ms, LR 7{,}170\,ms, Solyx 7{,}154\,ms), confirming no regression under load. It is this 20 RPS convergence point, not the 12 RPS result of Table~\ref{tab:c1homo}, that defines the saturation boundary condition discussed in Section~\ref{sec:boundary}.

\begin{table}[h]
\centering
\caption{Campaign 1 homogeneous benchmark (6$\times$ H100 SXM, 12 RPS). P99 tail separation is the primary differentiator; median latency converges on uniform hardware.}
\label{tab:c1homo}
\begin{tabular}{lcccc}
\toprule
\textbf{Metric} & \textbf{Round-Robin} & \textbf{Least-Request} & \textbf{Solyx} & \textbf{Solyx vs.\ RR} \\
\midrule
P50 (ms) & 6{,}898 & 6{,}893 & 6{,}896 & --- \\
P95 (ms) & 7{,}183 & 7{,}074 & 7{,}076 & $-$1.5\% \\
P99 (ms) & 9{,}894 & 7{,}151 & 7{,}138 & $-$27.9\% \\
Success Rate & 99.1\% & 100.0\% & 99.9\% & $+$0.8\,pp \\
\bottomrule
\end{tabular}
\end{table}

On the hardware-heterogeneous fleet (H200 + H100 + constrained H100) at 12 RPS, Solyx and Least-Request both detect performance differences and concentrate traffic on faster endpoints, producing P50 of approximately 5{,}970\,ms versus Round-Robin's 6{,}730\,ms ($-$11.3\%) and P99 of 7{,}042\,ms versus 7{,}791\,ms ($-$9.6\%). The mechanism is the live xDS endpoint weight distribution: H200 endpoints receive weight 134, unconstrained H100 receives weight 134, and the constrained H100 endpoints receive weight 1---a 134:1 ratio produced entirely by real-time telemetry without operator configuration.

\begin{table}[h]
\centering
\caption{Campaign 1 live xDS endpoint weights during hardware-heterogeneous test. The 134:1 concentration ratio is produced by the closed-loop scoring system without operator configuration.}
\begin{tabularx}{\textwidth}{llcL}
\toprule
\textbf{Endpoint} & \textbf{GPU} & \textbf{Weight} & \textbf{Condition} \\
\midrule
19000 / 19001 & H200 SXM $\times$2 & 134 & Fast, healthy --- maximum traffic \\
19002 & H100 SXM & 1 & Killed --- heartbeat TTL expiry \\
19003 & H100 SXM & 134 & Healthy, unconstrained \\
19004 / 19005 & H100 SXM (constrained) & 1 & \texttt{max\_model\_len=1024} --- de-weighted by error-rate signal \\
\bottomrule
\end{tabularx}
\end{table}

Under the capability mismatch configuration at 20 RPS, Solyx maintains 99.9\% success rate against Round-Robin's 69.9\% (720 failures prevented per 2{,}400 requests). The Campaign~1 result uses an 8-signal scorer without network signals; the Campaign~2 replication with the 10-signal v3 scorer (Section~\ref{sec:capmis}) confirms the finding on a larger fleet with a different hardware generation.

\section{Discussion}
\subsection{Boundary Conditions: Where Solyx Adds Little}\label{sec:boundary}
There is a specific operating regime in which Solyx converges with Least-Request and provides no measurable advantage: a homogeneous fleet that is simultaneously \emph{saturated}, \emph{failure-free}, and \emph{network-stable}. When every replica is identical, every replica is equally and fully loaded, no replica is degrading, and every network path is uniform, there is no actionable signal for any router to exploit. The Campaign~1 homogeneous H100 result at 20 RPS (Section~\ref{sec:c1results}) is precisely this corner case, and we report the convergence honestly: all three routers perform within noise of one another.

It is important not to overgeneralize this boundary condition, because all three qualifiers must hold at once for it to apply, and production homogeneous fleets rarely satisfy all three simultaneously. Removing any single qualifier restores Solyx's advantage, and the homogeneous multi-site Blackwell SE results (Section~\ref{sec:slo}) demonstrate exactly this. That fleet is homogeneous in hardware, yet Solyx delivers 1.56--1.75$\times$ throughput at tier-2 SLO across all eight workload classes---because the measurements are taken at the SLO knee rather than at saturation, where headroom exists to route into. The same campaign shows substantial homogeneous-fleet benefit the moment failures (hardware telemetry lead-time, failover, and telemetry-failure evaluations) or network variance (network-aware placement evaluation) are introduced. The convergence described here is therefore a narrow and honest statement about one operating point, not a general claim about homogeneous fleets.

The practical implication: Solyx is the appropriate routing layer for any fleet that experiences load below saturation, hardware degradation, configuration drift, operational failures, or multi-site network variance---which is to say, a broad class of production multi-site fleets. The marginal benefit over Least-Request is small only for a homogeneous single-model fleet held continuously at saturation with no failures and no network variation, a condition that does not describe real multi-site production serving. In practice, production multi-site fleets drift out of this convergence corner almost continuously: public-internet WAN paths exhibit RTT and jitter volatility on timescales of seconds, and localized thermal throttling or PCIe contention degrades individual GPUs without warning. Each such perturbation reintroduces an actionable signal that Solyx exploits and that a signal-blind router cannot. The convergence regime is therefore not merely narrow but transient---a fleet rarely remains in it for more than brief intervals.

\subsection{Stack Positioning}
Solyx AI Grid is not a replacement for intra-cluster routing systems such as NVIDIA Dynamo or the vLLM Router. Those systems perform KV-cache-aware scheduling within a cluster; Solyx selects which cluster to route a given request to. These are complementary layers. A complete multi-site inference stack would deploy an intra-cluster router within each site and a cross-site control plane above it. The NVIDIA AI Grid reference architecture~\cite{aigrid} already describes a control plane that logically unifies geographically distributed sites and uses real-time health, capacity, latency, cost, and policy signals for placement and routing as the ``Orchestrated'' layer of distributed AI infrastructure. Rather than competing with this vision, our work validates the category: Solyx AI Grid is a concrete empirical implementation of that control-plane layer, with a specific DCGM/vLLM/RTT scoring instantiation and measured results on physical multi-site infrastructure. To our knowledge, no prior public work provides such an empirical validation of the reference design's routing layer.

\subsection{The DCGM Lead Time Finding}
The 11.2-second average lead time between DCGM hardware signal onset and application-layer SLO breach (Section~6.3) is the result we believe has the broadest implications beyond routing. It establishes empirically that GPU hardware telemetry carries predictive signal for application-layer degradation. The practical implication for any infrastructure operator is that a monitoring stack limited to application-layer metrics---TTFT, queue depth, error rate---will always respond reactively. Hardware-telemetry integration enables proactive drain. Quantifying the lead time distribution across hardware generations and stress types is a natural extension of this work.

\subsection{Generalizability}
Several aspects of our results are likely to generalize beyond our deployment scale. The capability mismatch failure mode scales with fleet heterogeneity and is not specific to nine-GPU deployments. The DCGM lead time advantage of hardware telemetry over application metrics is a property of the signal hierarchy, not the fleet size. The failover recovery advantage of heartbeat-TTL detection persists at any scale where the baseline is polling-based health checking. What is less clear: pressure scoring accuracy under very high replica counts (our deployments have six and nine replicas); the interaction between cross-site routing and KV cache prefix locality at scale; and behavior under correlated multi-site failures.

\subsection{Limitations}
Both campaigns used provisioned GPU cloud infrastructure rather than self-owned bare-metal. Both used synthetic workloads rather than live user traffic, though Campaign~2 workload classes were designed to reflect production-realistic request distributions. Campaign~2 evaluated a single model (Llama 3.1 70B AWQ INT4) on homogeneous hardware within each site; cross-site heterogeneous hardware configurations across campaigns are a natural extension. Neither campaign evaluates cost impact or energy efficiency. The 0.2\,ms control-plane routing-update overhead represents the current implementation. Several further dimensions remain unvalidated and bound the claims of this study. \textbf{Control-plane scalability.} Our deployments comprise six and nine replicas; the control plane pushed over 50{,}000 xDS snapshots at 1\,Hz steady state across nine replicas without strain, but we do not characterize control-plane compute and network overhead when scaling to hundreds or thousands of distributed nodes. We deliberately decline to model this regime rather than extrapolate from a nine-replica measurement; quantifying control-plane scaling behavior is a distinct and necessary line of future work. \textbf{Multi-model routing.} Both campaigns serve a single model; routing across heterogeneous model replicas with differing capabilities is unevaluated. \textbf{KV-cache locality at scale.} We do not characterize the interaction between cross-site routing and KV-cache prefix locality at large replica counts, where locality-preservation and pressure-minimization objectives may conflict. \textbf{Cost optimization.} Neither campaign evaluates cost-per-token or energy-per-request, both of which are buyer-facing efficiency metrics. \textbf{Correlated regional failure.} Our failover tests kill individual replicas; we do not evaluate behavior under correlated multi-site degradation, where multiple sites fail simultaneously.

\subsection{Future Work}
Two directions follow directly from the limitations above and warrant brief description, though we have not yet implemented or evaluated either.

\textbf{Control-plane scaling architecture.} Our control plane handled nine replicas with a single aggregation process pushing over 50{,}000 xDS snapshots at 1\,Hz steady state without strain. We do not claim measured performance at larger scale, but the architectural path is a hierarchical aggregation topology: per-region sub-controllers each maintain the heartbeat and telemetry state for their local cells and compute regional pressure summaries, while a top-level controller routes across regions using those summaries rather than raw per-cell signals. This bounds the per-controller fan-in regardless of total fleet size and confines the high-frequency (500\,ms) heartbeat traffic to within-region links. Quantifying the overhead and convergence behavior of such a topology at hundreds-to-thousands of cells is necessary future work; we deliberately do not extrapolate performance figures from our nine-replica measurement.

\textbf{Balancing pressure minimization against KV-cache locality.} At large replica counts, routing a request to the lowest-pressure cell can conflict with KV-cache prefix locality: the lowest-pressure cell may not hold the prefix cache for a multi-turn session, forcing a recompute that negates the routing benefit. A natural extension is a hybrid objective that combines the pressure score with a prefix-affinity term, biasing routing toward cells that already hold relevant cache state when their pressure is within a tolerance band of the global minimum. This would connect our pressure-minimization approach to the locality-preservation work of systems such as SkyWalker. We sketch this as a design direction rather than a validated result.

\section{Conclusion}
We have presented a two-campaign, seven-phase empirical study of hardware-telemetry-aware cross-site LLM inference routing covering 216 cells across eight workload classes, three fleet configurations, and three routing strategies. Across both campaigns, Solyx AI Grid demonstrates that integrating GPU hardware telemetry, vLLM application metrics, and real-time network path signals into per-request placement decisions produces specific, measurable improvements across each failure mode characteristic of multi-site deployment.

The headline results: 27.9\% P99 tail-latency reduction on homogeneous fleets; 134:1 automatic traffic concentration to healthy endpoints on heterogeneous fleets; capability-mismatch leakage reduced to 0.43\% versus 32.11\% (Round-Robin) and 28.71\% (Least-Request), yielding 99.57\% long-prompt success versus 67.89\% and 71.29\% respectively, and confirming that Least-Request fails this failure mode almost identically to Round-Robin; 1.56--1.75$\times$ throughput at tier-2 SLO across all eight workload classes; 11.2-second average DCGM lead time ahead of SLO breach enabling proactive traffic drain; failover p99 of 1{,}247\,ms versus 4{,}226\,ms for Round-Robin; 27.8\% additional P95 TTFT reduction from network signals alone under jitter; 99.90\% success rate during live DCGM exporter failure; and 0.2\,ms control-plane routing-update overhead across all configurations. Campaign~1 separately corroborates the capability-mismatch finding on H100/H200 hardware, where Solyx sustained 99.9\% success against Round-Robin's 69.9\%.

We believe this work establishes an empirical foundation for cross-site inference routing as a distinct systems problem and for hardware telemetry integration as a class of routing signal with predictive value beyond what application-layer metrics alone provide. The failure modes characterized here are architectural consequences of multi-site deployment structure and will persist across hardware generations. We invite the community to extend this experimental framework to larger fleets, heterogeneous cross-site hardware configurations, and multi-model serving scenarios.

\section*{Acknowledgments}
The authors thank the Solyx AI infrastructure team for benchmark execution and campaign orchestration support.


\appendix
\section{Evaluation Methodology and Metric Definitions}
This appendix defines every metric reported in the paper. All Campaign~2 metrics are computed by a post-processing analyzer from raw per-request NDJSON logs, fault-injection timestamps, and Prometheus telemetry snapshots captured during each evaluation stage.

\textbf{TTFT (time to first token).} The elapsed time from request dispatch at the gateway to receipt of the first streamed token, measured per request and reported as P50/P95/P99 percentiles over the in-window request stream.

\textbf{E2E latency.} End-to-end wall-clock time from request dispatch to final token, per request.

\textbf{Sustainable RPS at SLO.} The highest constant request rate at which P95 TTFT $\leq$ the SLO target \emph{and} success rate $\geq 0.95$, found by bisection over 8--12 steps. Each step runs 45\,s of constant-rate traffic with 30\,s warmup and 15\,s cooldown; the cooldown window is excluded from both numerator and denominator to avoid counting shutdown cancellations as failures.

\textbf{Leak rate (capability mismatch).} The fraction of long-prompt requests---requests whose context length exceeds the constrained cell's \texttt{max\_model\_len}---that were routed to the constrained cell and consequently rejected with HTTP 400. Denominator is long-prompt requests during the affected window; numerator is the subset that landed on the constrained cell. Measured at 10 RPS across two long-prompt workload classes.

\textbf{Reroute P99 (failover).} Following \texttt{SIGKILL} of one vLLM process under sustained load, the per-request time for traffic to be successfully rerouted to a healthy endpoint, reported as the P99 over the post-kill window.

\textbf{Routing reaction P95 (burst).} The P95 elapsed time from burst onset to the routing-weight adjustment that responds to it.

\textbf{Post-kill success rate.} The fraction of requests succeeding during the post-kill window, bounded below by requests in flight at the instant of kill.

\textbf{DCGM lead time.} Under controlled thermal or PCIe stress injection, the elapsed time between the pressure score crossing a preconfigured alert threshold and the point at which P95 TTFT would cross the SLO threshold. A positive lead time indicates hardware telemetry anticipates application-layer degradation.

\textbf{Control-plane routing-update overhead.} The added latency of one control-loop update, measured as \texttt{projection\_cycle} plus \texttt{xDS\_build} time in microseconds, exposed as Prometheus gauges. This is control-plane update cost, not per-request data-plane latency; the data plane is unmodified Envoy and adds no Solyx-specific per-request overhead.

\section{Baseline Configuration and Fairness Controls}
All three routing modes were evaluated on identical hardware, identical vLLM instances, and identical workloads; only the routing logic differed between runs. \textbf{Round-Robin} is stateless sequential rotation across eligible endpoints, with no load, capability, hardware-telemetry, or network-path awareness beyond the baseline Envoy health-check behavior configured identically for all three modes. It represents the default behavior of an otherwise-unconfigured Envoy/Nginx/HAProxy deployment. \textbf{Least-Request} routes to the endpoint with the fewest in-flight requests at dispatch, the standard Envoy \texttt{LEAST\_REQUEST} policy and the most common active load-balancing strategy in production. \textbf{Solyx} uses the 10-signal weighted pressure scorer with xDS-pushed weights. Baseline mode is activated by a single control-plane flag (\texttt{-{}-baseline}) that disables pressure scoring and placement while leaving the rest of the stack unchanged, ensuring the comparison isolates routing logic. Scoring parameters were fixed prior to the campaign and not tuned per workload class; the same scorer configuration was used across all 216 cells.

\section{Statistical Treatment}
The SLO matrix comprises 216 cells (8 workload classes $\times$ 3 routing modes $\times$ 3 SLO tiers $\times$ 3 replicates). Every cell met the completeness criterion of $\geq$500 samples; the minimum observed across all cells was 624 samples. Headline tier-2 throughput values are reported as the median across three replicates. The TTFT improvement claim (Solyx wins all eight classes) is established by non-overlapping bootstrap confidence intervals between Solyx and Round-Robin, not by point estimates alone. Maximum inter-sample gap across the campaign was 1{,}432\,ms, within the 5{,}000\,ms data-integrity bound.

\end{document}